\documentclass[12pt]{elsart}
\usepackage{graphicx}
\begin{document}
\baselineskip 20pt plus .1pt  minus .1pt \pagestyle{plain}
\voffset +1.0cm \hoffset 0.0cm \setcounter{page}{01} \rightline
{} \vskip 1.0cm

\begin{center}
\Large {ON THE CORRELATION BETWEEN COSMIC RAY INTENSITY AND CLOUD COVER}\\
\end{center}
\vspace {0.5cm}  
\begin{center}
\footnote{E-mail address: erlykin@sci.lebedev.ru} 
A.D.Erlykin$^{(1,2)}$, G.Gyalai$^{(3)}$, K.Kudela$^{(3)}$, T.Sloan$^{(4)}$, A.W.Wolfendale$^{(2)}$
 
\end{center}
\begin{flushleft}
(1) {\em P. N. Lebedev Physical Institute, Moscow, Russia}\\
(2) {\em Dept. of Physics, Durham University, Durham, UK}\\
(3) {\em Inst. Exp. Phys. Slovak Acad. Sci., Kosice, Slovakia}\\
(4) {\em Dept. of Physics, Lancaster University, Lancaster, UK} 
\end{flushleft}

\begin{abstract}
Various aspects of the connection between cloud cover (CC) and cosmic rays (CR)
are analysed. Most features of this connection viz. an altitude dependence of the 
absolute values of CC and CR intensity, no evidence for the correlation 
between the ionization of the atmosphere and cloudiness, the absence of  
correlations in short-term low cloud cover (LCC) and CR variations indicate that there 
is no direct causal connection between LCC and CR in spite of the evident long-term 
correlation between them. However, these arguments are indirect. If only some 
part of the LCC is connected and varies with CR, then its value, obtained from 
the joint analysis of their 11-year variations and averaged over the Globe, should be 
most likely less than 20\%. 

The most significant argument against causal connection of CR and LCC is the 
anticorrelation between LCC and the medium cloud cover (MCC). The scenario of the 
parallel influence of the solar activity on the Global temperature and CC from one side
and CR from the other side, which can lead to the observed correlations, is discussed 
and advocated.   
\end{abstract}
\section{Introduction}
The correlation between CR and LCC variations, which led to the introduction of a 
new scientific subject - 'cosmoclimatology', was found more than 10 years ago 
(~Svensmark and Friis-Christensen, 1997; Palle Bago and Butler, 2000; Svensmark, 2007~)
. The proponents of the causal connection between CR and LCC point out a number of 
facts. Firstly, there is the positive character of the correlation, i.
e. an increase of the CR intensity is accompanied by an increase of LCC and vice versa.
 Secondly, the peak to peak 
amplitude of the Global LCC variations ($\sim$2\%) is much higher than the amplitude
of the variation of the energy flux, delivered by the Sun, or the sun's irradiance 
(SI,$\sim$0.1\% peak to peak over the Solar Cycle), which requires us to find the 
mechanism which would explain such a large magnification.
Thirdly, there is an effect in the LCC similar to the latitude effect in CR, 
i.e. LCC variations during the 11-year cycle of solar activity are less in the tropics
than at higher latitudes. The same reduction of variations in the equatorial regions 
exists also in CR due to the higher geomagnetic rigidity near the equator.

The opponents of the causal connection between CR and LCC have put forward different 
arguments. Firstly, the positive correlation of CR and Cloud Cover (CC) is noticed only
 for the LCC, 
i.e. for clouds below 3 km above sea level. No significant positive correlation has 
been found for higher clouds. Secondly, there is an altitude dependence of CC and CR, 
but it changes sign. 
If one thinks about the ionization of the air as the mechanism of the CR influence on 
CC formation (~the usual assumption~), 
then the maximum of the CR flux and ionization is at heights of $\sim$12-15 km and not 
below 3 km, where the effect is claimed. Thirdly, there were no changes of CC noticed 
after the 
significant release of radioactivity during the Chernobyl disaster or during 
ground-based tests of nuclear weapons (~Erlykin et al., 2009~). Fourthly, there 
were no CC changes found during and after strong short-term variations of the CR 
intensity (~Forbush decreases or GLE - ground level events~) (~Kristj\'{a}nsson et al.,
2008, Sloan and Wolfendale, 2008~). 

The purpose of the present paper is a further analysis of the possible origin of LCC 
and CR correlations found in the work of Svensmark and Friis-Christensen, (1997) and 
Palle Bago and Butler, (2000).
 
\section{Input data}
As input data on the CC we used the same observations by meteo-satellites
 incorporated in the ISCCP program (~ISCCP, 1996~), which were used in 
Svensmark and Friis-Christensen, (1997); Palle Bago and Butler, (2000); Svensmark, 
(2007). We analysed 
monthly means for the fraction of the total observed area occupied by the clouds (D2).
 Following the classification of the cloud heights adopted in the ISCCP they were 
classified according to the pressure at their top border as: low ( LCC, $>$680 hPa ), 
medium ( MCC, 440-680 hPa ) and high ( HCC, $<$440 hPa ). Due to the continuing dispute
 on the quality of ISCCP radiometer calibrations after 1996 (~Marsh and Svensmark, 
2003~) we started the analysis using data obtained only during the 22nd cycle of the 
solar activity ( July 1986 - December 1995 ), but later added also the 23rd cycle and 
used the whole set of data available from then on. For 
the comparison with CR variations we used as a proxy of the Global CR intensity just
the neutron counting rate of the Climax neutron monitor, situated at a latitude of 
$39.4^\circ N$ (~WDC neutron data~). The CR variations at other latitudes, though 
having different amplitudes, have the same temporal behavior. The differences in 
amplitudes of the variations do not influence the value of the correlation coefficient.
 
In the analysis of the latitude dependence of CR and CC variations the entire latitude 
range from -90$^\circ$ to 90$^\circ$ was divided into 9 equal intervals of  
20$^\circ$ width. We analysed also the temporal behavior of the Global CC, i.e. 
averaged over the Globe. For the more distinct revelation of the non-trivial variations
 of CC in most cases we subtracted seasonal variations of CC from winter to 
summer, but in special cases we analysed also total variations including seasonal ones.
Seasonal CC variations were calculated as deviations of the monthly mean CC values in 
the D2 series from the yearly mean values averaged over all similar months (~January 
through December~) used in the analysis.
\section{Results}
\subsection{The altitude dependence of the Cloud Cover}
The mean values of the Global CC during the 22nd solar cycle are (28.09$\pm$1.06)\% for
 LCC, (19.52$\pm$1.70)\% for MCC and (13.35$\pm$0.61)\% for HCC. One can see that CC 
goes down with increasing altitude, which is opposite to the rising behavior of CR, 
 see e.g. Hayakawa, (1965). Numerous models have been proposed to explain this 
different altitude dependence and justify the causal CR-CC  connection (~see eg. the 
bibliography in Kirkby, 2007 or the recent paper by Kudryavtsev and Yungner, 2009~), 
but in our view there is still no convincing proof of their validity. The 
different altitude dependence of CC and CR is a problem for the concept of causal 
connection between them and requires further study. 

The most likely part
of CR which can be connected with cloud formation is their charged component, which
produces ionization and which could in principle give rise to the growth of 
condensation nuclei. Balloon 
studies of temporal variations of the charged CR component at different atmospheric 
altitudes show that the correlation between variations of the charged particle flux and
 the counting rate of ground-based neutron monitors, which is rather high in the 
stratosphere above 15 km, decreases below 6 km (~Bazi1evskaya et al., 2007; 
Bazilevskaya et al., 2008; Ermakov et al., 1997~). The correlation coefficient at 
altitudes below 3 km becomes as low as $\sim0.2$. So it is hard to expect that CR 
variations observed with neutron monitors could be the cause of LCC variations via  
ion production.   
\subsection{The time lag between CR and LCC temporal variations}
In Figure 1 the temporal behavior of the CR intensity (a) and Global LCC (b) are shown 
for solar cycle 22 ( 1986 - 1996 ). 

Figure 1.

As an illustration of the CR 
behavior we have taken the data of the Climax neutron monitor. The 
qualitative correlation between CR and LCC can be seen by the naked eye: both of them 
reach their minimum at about the same time - July-October 1990. 
One can ask whether it is possible to find the best fit time lag between 
these curves, for which the least-squares $\chi^2$ between them has a  
minimum. One can imagine that if, say, CR variations start after the LCC ones, 
then CR can hardly be the cause of the LCC variations. The value of $\chi^2$ was 
calculated as 
\begin{equation}
\chi^2(\Delta t)=\Sigma_{i=1}^{ndf}(\frac{LCC}{\langle LCC \rangle}(t_i+\Delta t)-\frac{I_{CR}}{\langle I_{CR} \rangle}(t_i))^2
\end{equation}
Here $LCC(t_i+\Delta t)$ is the Global low cloud cover value at the time $t_i+\Delta t$
, $\langle LCC \rangle$ is its mean value in the studied time interval 1986-1996, 
$I_{CR}(t_i)$ and $\langle I_{CR} \rangle$ are the CR intensity and its mean value 
respectively. $ndf$(number of 
degrees of freedom) is the number of months taken in the analysis, $\Delta t$ is the 
time lag between $LCC$ and $I_{CR}$, for which we find the minimum of $\chi^2$.   
  
In Figure 1c we show the value of $\chi^2/ndf$ as a function of this
time lag within a $\Delta t = \pm1$ year time interval. 
It is seen that $\chi^2/ndf$ has a very flat and broad 
minimum within -11/+6 months time lag and it is not possible to say which of CR and LCC
 variations start first.  
\subsection{Long-term variations and the fraction of LCC which correlates with CR}
As {\em long-term variations} we call deviations of LCC and CR values from their 
means obtained by averaging over the entire analysed time interval (~the dotted lines 
in Figures 1a,b~). In Figure 2a we show the correlation plot for the variations of LCC 
and CR during the 22nd solar cycle.

Figure 2

It is seen that deviations from the mean values of LCC and CR correlate positively 
with each other. The slope of the linear regression line is 0.157$\pm$0.023 and the 
correlation coefficient is 0.538$\pm$0.047, which confirms the positive correlation 
between LCC and CR, found in Svensmark and Friis-Christensen (1997), Palle Bago and 
Butler (2000). 

If it is assumed that CR are responsible for just a fraction of the LCC and they are 
the \textbf{only} agent creating this fraction, then from the 
observed correlation it is possible to estimate how large this fraction is. Such an 
estimate depends on the model of the connection between CR and LCC. Let us assume that
this connection can be fitted as
\begin{equation}
Y=a+bX^c 
\end{equation}
where $Y=\frac{LCC}{\langle LCC \rangle}$ and $X=\frac{I_{CR}}{\langle I_{CR} \rangle}$
. Here $LCC$ and $I_{CR}$ are values of LCC and CR intensity and $\langle LCC \rangle$ 
and $\langle I_{CR} \rangle$ are their mean values respectively. The first and second 
terms in this
expression determine the parts of LCC independent and dependent on CR respectively.
 If the connection between CR and LCC is linear, i.e. $c = 1$, then the slope of the 
linear regression line, $b=0.157$, gives the fraction of LCC connected with CR as
  $\sim$16\% as the best estimate and which should not exceed 20\% at the level of 2 
standard deviations. However,
 the determination of the $a,b,c$ coefficients by the least-squares method shows that 
the best-fit connection between CR and LCC is non-linear rather than linear. The 
derived values are $a=0.9783\pm 0.0008$, $b=0.0177\pm 0.0010$ and $c=8.65\pm 0.46$ 
(~full line in Figure 2a~). This shows that the most likely fraction of LCC connected 
with CR, which can be derived from expression (3), does not exceed 2\% around $X=1$. 

This conclusion is valid only if the models of the CR and LCC connection are true 
and CR variations at the Climax latitude of $\sim$40$^\circ$N are a good representation
  of the Global CR variations. For values of $c < 1$ the fraction of LCC, which varies 
together with CR, can be higher. Unfortunately, due to the relatively small magnitude 
of the CR and LCC variations, it is impossible to distinguish between the models of the
 connection from the LCC-CR correlation plot of Figure 2a. Although the least-squares  
method gives preference to the value of $c>1$, in the region where there are 
experimental data the behavior of curves for different values of $c$ and corresponding
 least-squares sums differ insignificantly from each other.

The two most popular models adopted for the connection between LCC and CR, which are 
discussed in the literature, will be considered. They are based on the connection 
between the ionization rate $q$ and ion density $n$ in the atmosphere. The first model 
assumes that $n \propto \sqrt{q}$, the second one - $n \propto q$ (~Mason, 1971; 
Bazilevskaya et al., 2008~). 
Applied to the LCC-CR connection they correspond to $c = 0.5$ for the first 
model and $c = 1$ for the second one. It is appreciated that elsewhere (~Sloan and 
Wolfendale, 2008~) 
the $n \propto \sqrt{q}$ model was used. Were that to be adopted here the upper limit 
to the CR fraction would go up by $\sim$ factor 2, to 40\%. Conversely, if 
$n \propto q$, the Sloan and Wolfendale limit (~Sloan and Wolfendale, 2008~) 
would fall to 12\% at the 95\% confidence level. 

Experimental data for the charged CR and ion density, which is, of course, relevant 
here, give preference to $c = 1$ and show no evidence for a change with altitude at
 least for altitudes about 7 - 30 km above sea level. At lower altitudes they indicate
the trend to $c > 1$, which qualitatively agrees with our best fit value of $c = 8.65$
(~Ermakov et al., 1997~). Keeping in mind all the necessary 
reservations we persist with 
 our estimate of the fraction $f < 20\%$ for the latter model since it is based on 
the experimental data.  

The authors referred to above (~Bazilevskaya et al., 2007~) also stressed that 
variations observed with ground-based neutron monitors correlate well with charged CR 
fluxes only at altitudes above 15 km, thus, they may be correctly used as a proxy of 
ionizing component only for stratospheric 
altitudes. At altitudes below 3 km the correlation coefficient falls to about 0.2.
Therefore it is unlikely that variations observed with neutron monitors can be followed
 by similar variations of the ionizing component at low altitudes. The good positive  
correlation between the counting rate in neutron monitors and LCC found in 
 Svensmark and Friis-Christensen (1997); Palle Bago and Butler (2000) should have a  
cause different from the ionization of the air 
 by CR with subsequent formation of cloud droplets on these ions.   
\subsection{Short-term variations}
Figure 2a and the analysis made in the previous subsection are relevant to the total CR
 and LCC variations about their mean value, the main contribution to which is given by
the long-term variations, connected with the 11-year cycle of solar activity. In 
order 
to reveal the possible correlation of {\em short-term CR and LCC variations} we removed
 the contribution of long-term variations. For that purpose the temporal behavior of CR
 and LCC were approximated by a 5-degree polynomial fit (~dashed lines in Figures
 1a and 1b~) and deviations from this fit were calculated. Since we used the D2-set, 
i.e. 
monthly averaged data, this analysis relates to the variations of monthly duration.
We did not find any significant correlation between CR and Global LCC (~Figure 2b~).
 The slope of the linear regression line was $b=-0.060\pm0.062$ and the correlation 
coefficient $r = -0.104\pm0.092$. This negative result is in indirect agreement with 
the absence of even shorter daily-long variations of LCC during Forbush decreases or 
GLE, analysed in Kristj\'{a}nsson et al. (2008); Sloan and Wolfendale (2008).

The preliminary conclusion which can be drawn from the analysis so far is the 
following: if CR are resposible for a part of the LCC then it is most likely that 
this part is small, viz. less than about 20\%. The absence of short-term correlations 
between CR and LCC indicates that the assumed causal connection between them could be 
revealed only on a longer time scale, not less than several months, which could be 
understood if the Global LCC has a monthly or longer inertia.      
\subsection{The anticorrelation between LCC and CC at higher altitudes: MCC and HCC}
A significant argument against the causal connection between CR and LCC is the 
anticorrelation of LCC and CC at higher altitudes: MCC and HCC. In Svensmark and 
Friis-Christensen (1997); Palle Bago and Butler (2000)
 the authors claim 
that they cannot find any positive correlation between CR, MCC and HCC, similar to that
 found for LCC. It is true, since both MCC and HCC {\em anticorrelate} with LCC. This 
is illustrated in Figure 3 for the same 22nd solar cycle.     

Figure 3

The left set of panels shows the temporal behavior of the Global MCC and LCC together 
with the correlation between their variations. The anticorrelation is clearly seen both
 in long-term and short-term variations. The existence of short-term anticorrelations
between MCC and LCC is a remarkable difference with the case of CR and LCC. It 
points to a strong connection between clouds at adjacent altitudes.

The right hand set of panels shows the same characteristics for clouds at the adjacent 
altitudes: HCC and MCC. The positive long-term correlation between them proves the 
existence of an anticorrelation between HCC and LCC. Short-term correlations between
HCC and MCC are absent. A relevant point concerns the role of updrafts which play such 
a key role in cloud formation; they can in principle cause the LCC and MCC 
anticorrelations (~see later~).

Turning to HCC with mean temperatures below $\sim -30^\circ C$ at all latitudes, ice 
crystals are important and the Physics is different. The lack of an HCC-MCC 
correlation is not surprising. 

The anticorrelation between LCC and CC at higher altitudes gives a strong argument 
against the causal connection between CR and LCC. It is difficult to imagine that, 
say, the rise of CR intensity could raise LCC below 3 km, but reduce MCC above this 
altitude and vice versa.  
\subsection{Seasonal variations}
In all previous figures well understood seasonal variations of CC have been removed. 
However, they can also be used to clarify the interaction of clouds at different
atmospheric altitudes. Figure 4 shows the temporal behavior of MCC and LCC during the 
last decade of the century.
It is clearly seen that the minima of MCC correspond to maxima of LCC and vice versa.
Therefore, the anticorrelation of long-term decadal and short-term monthly variations 
of MCC and LCC, illustrated in previous subsections, is strongly confirmed on the 
intermediate yearly time scale. 

Figure 4.

It has been argued that the strong seasonal periodicity of CC is most likely caused by 
the seasonal variation of the surface temperature $T$. Despite the fact that the 
seasons are opposite in the northern and southern hemispheres the surface temperature 
averaged over the Globe still depends on the season. The amplitude of the variation is 
about $3.8^\circ C$ (~Figure 5~) and its maximum is in July (~Global Surface 
Temperature Anomalies~).

Figure 5.

Figure 4 shows that the maximum of Global LCC is also in the middle of the year. 
If LCC and MCC variations are directly connected with variations of the Global surface 
temperature it means that either the LCC-$T$ correlation is positive and there is no
time lag between them longer than 2-3 months. If the thermal inertia of the Earth's 
surface causes longer time lags of about 6 months or longer the LCC-$T$ correlation 
should be negative. The latter possibility seems to us more likely since it is 
confirmed by the existence of the time lags between $T$ and LCC long-term variations.
(~see later~). The possible mechanism of the $T$ and CC connection should also give
an opposite sign for LCC and MCC variations.
\subsection{CR and CC correlations in the 22nd and 23rd solar cycles}
At the beginning of the 21st century it was noticed that the positive correlation
between CR and LCC (~Marsh and Svensmark, 2003; Usoskin et al., 
2004~) decreased. While the minimum CR intensity at the 
23rd solar cycle around 2001-2003 increased compared with the previous minimum in 
1990-1992, the LCC in the 23rd solar cycle was definitely lower than in the 22nd cycle 
(~Figure 6~).

Figure 6.

It should be added that many studies have found cycle 23 to be 'anomalous' in a number 
of ways: lower sunspot number, but nearly no change in total solar irradiance, extended
 solar minimum, a hump in the neutron monitor counting rate in 2004 and 2005 etc.

 When we include the 23rd cycle into our analysis the correlation coefficient falls 
from $r=0.538$ down to $r=0.390$. Some authors (~Marsh and Svensmark, 2003; Usoskin 
et al., 2004~) explained this fact by a fault in the 
calibration of the ISCCP radiometers, which occured at about 1995, and caused the 
continuous decreasing trend in the derived LCC values.
We think, however, that this trend is {\bf not} an artifact connected with that fault, 
but is {\em a real physical effect connected with the rising surface temperature}. In 
Figure 7 the temporal behavior of the Global LCC, MCC and HCC are shown together with 
the rising temperature during the 22nd and 23rd cycles of solar activity.

Figure 7.

With the addition of the 23rd solar cycle, the anticorrelation of LCC with MCC and HCC
becomes even stronger. The long-term anticorrelation between LCC and MCC increases from
$r = -0.636$ up to $r = -0.873$, the short-term anticorrelation - from $r = -0.510$ up 
to $r = -0.585$. We argue that the strong short-term and seasonal anticorrelations 
between LCC 
and MCC are a serious argument against an artificial origin of the long-term decrease
of LCC. The assumed calibration fault of ISCCP radiometers can hardly have a monthly
occurence and seasonal periodicity. In what follows we shall analyse the CC behavior
mostly taking into account the 22nd and 23rd solar cycles together.
\subsection{The latitude dependence of CC properties}
Both the CR intensity and the surface temperature depend on the latitude. In this 
connection it is also reasonable to analyse the variation of the CC characteristics 
with latitude. We show some of them in Figure 8.

Figure 8.

Figure 8a shows the latitude dependence of LCC, MCC and HCC. It is seen that there is a
 small minimum for LCC in the equatorial region, which could in principle be connected 
with the reduction of the CR intensity, but it is not 
confirmed by the local maxima in MCC and HCC. In the Polar regions, where the
 CR intensity is highest, there is an opposite decrease of LCC, which apparently is
connected with the dominant influence of the atmospheric conditions, eg. low 
temperatures. The highest LCC is in the southern latitude bands with the largest part 
of the area occupied by oceans, i.e. with a relatively large density of water vapor. 

The altitude dependence of CC also does not correspond to the altitude dependence of 
the CR intensity. In most latitude bands MCC and HCC are smaller than LCC, which is
opposite to CR with their intensity rising with altitude. All this shows that even if
there is a causal connection between CR and LCC, its character is more complicated than
 the direct and positive connection.

We have already mentioned in \S3.3 that the Global LCC - CR correlation is  positive: 
$r=0.538$. Figure 8b shows the latitude dependence of the CC - CR correlation 
coefficient. For a CR proxy we used just the neutron counting rate at Climax. 
In spite of the latitude dependence of the CR variation amplitude, the value of the 
LCC -CR correlation coefficient does not depend on the latitude due to the similarity
of the temporal behavior of CR variations at different latitude bands. It is remarkable
 that, in most latitude bands, MCC and HCC have negative correlations with CR, in 
opposition to the positive LCC - CR correlation, which was the main argument for the 
claimed causal CR - CC connection (~Svensmark and Friis-Christensen, 1997; Palle Bago 
and Butler, 2000~). Furthermore, the general
 similarity of the MCC-CR and HCC-CR correlations does not fit in with the idea that 
CR-induced ions cause cloud droplets because in HCC ice crystals dominate, where, as 
remarked already, the Physics is different.

Figure 8c shows the latitude dependence of the sensitivity and correlation between
MCC and LCC. The sensitivity of one variable to another, according to the definition 
(~Uchaikin and Ryzhov, 1988~), is the derivative of the first variable on the second in
 log-coordinates. In our 
case the sensitivity is the slope of the linear regression line in the MCC-LCC plot. 
One can notice two features: (i) the sensitivity of MCC to LCC and MCC - LCC 
correlation coefficient are negative at nearly all latitudes, which is another support
 of their Global anticorrelation. The negative sensitivity of MCC to LCC is difficult 
to explain in the framework of the causal connection between CC and CR, since the rise 
of the CR intensity should change CC similarly at all altitudes ; 
(ii) the highest negative sensitivity and the correlation between MCC and LCC is 
observed in tropical and subtropical regions: $\ell = -30^\circ / +30^\circ$ as well as
 in the southern latitude bands with the highest fraction of water: 
$\ell = -65^\circ / -45^\circ$.
\section{Discussion} 
The vast bibliography of the works which have been devoted to the problem of the 
possible connecion between CR, CC  and climate is given in the comprehensive survey by 
Kirkby (2007).

The analysis made in the present work, as well as arguments presented in our previous 
publication (~Sloan and Wolfendale, 2008~), gives sufficient basis to argue that CR are
\textbf{not} the dominant factor in the formation of clouds. Long-term, short-term and 
seasonal anticorrelations of LCC and MCC, which are strongest in tropical and 
subtropical regions, as well as in regions mainly occupied by oceans, allow us
to return to the traditional scenario of the main cause of the cloud variation, 
connected with variations of the surface temperature, humidity and wind velocity. 
However, one can ask whether the temperature or CC variations start first. 

We try to answer this 
question analysing the time lag between the surface temperature {\em T} and LCC, since 
the lowest heights of the atmosphere are closest to the Earth's surface and the LCC is 
most sensitive to the surface temperature. For this purpose we come back to Figure 7 
which shows the temporal behavior of the Global surface temperature (panel {\em a}) and
 LCC (panel {\em d}) fitted by linear and 5-degree polynomial approximations.  

It is seen that besides its long-term rising trend the surface temperature has also an 
oscillating behavior similar to that of the LCC. Its amplitude is about 
$0.1^\circ C$ and the phase anticorrelates with the phase of LCC with a considerable
 time lag. It is difficult to estimate the magnitude of this time lag because it 
depends on the degree of the polynomial fit. A more detailed analysis of the 
correlation between LCC and surface temperature 
as a function of the time lag shows that the minimum negative correlation coefficient 
is for the time lag of 5 months, but the minimum is rather broad. Since temperature 
variations are ahead of LCC variations, one can conclude that the former could be the 
cause of the latter, but not vice versa. 

The long-term oscillations of temperature of the order $0.1^\circ C$ are observed in 
much longer time intervals (~Haigh, 2007; ACRIM~). They are usually associated with 
oscillations of the total solar irradiance (TSI) which has an 11-12 year periodicity. 
We have analysed the frequency spectrum of temperature variations for the 1880 - 2008 
time interval with the result shown in Figure 9.   

Figure 9.

One can notice the small peak at 0.007 month$^{-1}$ frequency, which corresponds to
$\sim$11-year period, coincident with the 11.87-year period of the solar cycle (~Sturrock,
2008~). 
The small amplitude of the peak and its corresponding low confidence level (~2.1 
standard deviations~) is presumably determined by the small amplitudes of TSI
 variations (~1.7 Wm$^{-2}$~) and of the corresponding solar forcing (~0.3 Wm$^{-2}$~)
together with the spread in '11 year' periods.
 We remark that the similar peak in the spectrum of land temperature variations is 
higher by a factor of 2, which is reasonable since the land is more sensitive to TSI
(~see Figure 5~). Interestingly, there is another peak in the frequence spectrum at 
0.004 month$^{-1}$, which corresponds to a 21-year solar cycle and has much higher 
confidence level (~7.1 standard deviations~). Therefore, periodic variations of the 
surface temperature and corresponding variations of LCC are most likely of solar or
perhaps geomagnetic origin rather than CR, because the 21-year variation in the CR rate
 is small.

 The most likely cause of the
 anticorrelations between LCC and MCC is the variation of convection flows of the air 
with temperature. The rise of the surface temperature gives rise to the growing 
temperature in the lower atmosphere and an upward convection flows with the 
corresponding rise of mean cloud heights. Since clouds in the ISCCP experiment are 
classified by the height of their upper borders the increase of their heights leads to 
the redistribution along their altitudes: some low clouds cross the 3 km border and 
become medium clouds. As a result, LCC decreases with rising temperature. This trend
is clearly seen in Figure 7, where the rise of the Global temperature in the (a)-panel 
is accompanied by the fall of LCC in panel (d).

The heating of the atmosphere is a slow process. The slow updraft of the air is the 
cause of the time lag between the variations of the surface temperature and CC. Perhaps
this time lag is the reason why the maximum of the seasonal Global temperature, which 
is in July (~Figure 5~), corresponds to the maximum in the seasonal variations of LCC, 
which is also in the middle of the year (~Figure 4~), in spite of their anticorrelation
in the long-term scale. The slow, but steady, upward convection flow of the heated air
is the possible mechanism for the magnification of small TSI variations ($\sim 0.1\%$) 
giving rise to larger variations of cloud heights ($\sim 0.7\%$). The slow fall of the
low cloud top pressure is seen in Figure 10. The slope of the linear fit $b$ is 
definitely negative: $b = -0.148\pm0.081$ hPa$\cdot$year$^{-1}$. It corresponds to a 
rise of the mean
low cloud top height by about 40m in 20 years. It is very slow but one should keep in 
mind that the observed {\em mean} cloud top height ($\sim$2.6km at 730hPa) is only 
$\sim$600m below the border ($\sim$3.2km at 680mPa) between LCC and MCC and any change 
in the height of clouds can cause the variation of the magnitude of LCC and MCC. 

We argue that the positive correlation of CR and LCC found in Svensmark and 
Friis-Christensen, (1997) and Palle Bago and Butler, (2000) is not evidence
for a causal connection between them, but the consequence of a parallel influence of
the common source - the solar activity on CR from one side and CC the other.

Concerning the relationship between CC and ground level temperature changes, there have
 been a number of studies for particular regions. Data for the USA covering the period 
1900 to 1990 (~Barry and Chorley, 1998~) give roughly a total 'mean annual cloud cover'
 change of 
-1.5\% over the 11-year cycle with an associated 0.2$^\circ C$ change in ground level 
temperature. A value for land higher than the Global average (~0.1\%~) is to be 
expected and there is no inconsistency with 'our' value.

Similar conclusions that most of the LCC variability comes from the subtropical oceans 
and is most likely due to TSI variations, causing changes in lower tropospheric static 
stability, have been made by Kristj\'{a}nsson et al. (2004).  

Another aspect of the Physics behind the correlation may be related to the relationship
between cloud height and cloud cover (~Cotton and Anthes, 1959~). These workers 
estimate that changes
 of -0.3\% in CC from a height of 5km (~mid MCC~) correspond to a change of 
+0.1$^\circ C$ at ground level (~i.e. cloud absorption of incoming radiation 
dominates~). In our case a change of +0.1$^\circ C$ at ground level corresponds to a 
change in LCC + MCC of -0.3\%, i.e. the same result although using the whole of MCC is 
not really appropriate. Nevertheless, there is a similarity in the values.

\section{Conclusion}  
We advocate a scenario for the origin of correlations between CR and LCC, based on 
the parallel influence of solar activity. 
 The solar irradiance rises with the sunspot number
in the middle of the solar cycle. The radiation is strongest in the tropics and 
subtropics. Though the relative rise of the irradiance is small, and only about 0.1\%, 
it causes a rise of the mean surface temperature and an increase of the vertical 
convection flows of the heated air. The subsequent change in supersaturation of the air
 at different heights can cause the changes in LCC and MCC. Warm air from below 3 km 
rising to greater heights will cause the LCC to fall and MCC to rise. By this way the 
rise of convection flows leads to a considerable magnification (to $\sim$2\%) of the 
effect of enhanced solar irradiance. Formulating briefly, one can say that in the 
maxima of the solar cycles the updraft becomes stronger and this effect is strongest in
 the tropics and subtropics, as well as in the southern latitude bands where
 there is the  largest fraction of area covered by the oceans. It is well known that 
the variations of solar activity are followed by the variations of CR intensity at 
Earth; the reduction of CR intensity coincident with the reduction of LCC is therefore 
by no means evidence of the causal connection between these two phenomena - they 
correlate with each other due to their common origin - the change of solar irradiance 
at the Earth. 

{\bf Acknowledgements}          
    
K.Kudela wishes to acknowledge the VEGA grant agency, project 2/7063/27. Erlykin A.D. 
expresses his deep gratitude to the John C. Taylor Charitable Foundation for financial 
support of this work, to the staff of Climax neutron monitor \\ 
$(http://ulysses.sr.unh.edu/NeutronMonitor/neutron_mon.html)$ \\and to E.V.Vashenyuk 
from Apatity neutron monitor \\$(http://pgi.kolasc.net.ru/CosmicRay)$ \\for access to 
their data. He thanks also S.P.Perov for useful discussions.           
\section{References}
\begin{itemize}
\item[] ACRIM: $http://acrim.com$
\item[] Barry, R.G. and Chorley, R.J., 1998, `Atmosphere, Weather and Climate', 
Routledge, London, New York.
\item[] Bazilevskaya, G.A., Krainev, M.B., Makhmutov, V.S., Svirzhevskaya, A.K., 
Svirzhevsky, N.S., Stozhkov, Yu.I., 2007, Variations of Charged Particle Fluxes in the 
Earth's Troposphere, Bulletin of the Lebedev Physics Institute, $\bf{34}$, 348.
\item[] Bazilevskaya, G.A., Usoskin, I.G., Fl\"{u}ckiger, E.O., Harrison, R.G.,
Desorgher, L., B\"{u}tikofer, R., Krainev, M.B., Makhmutov, V.S., Stozhkov, Yu.I.,
Svirzhevskaya, A.K., Svirzhevsky, N.S. and Kovaltsov, G.A., 2008, Cosmic Ray Induced 
Ion Production in the Atmosphere, Space Science Reviews, $\bf{137}$, 149.
\item[] Erlykin, A.D., Gyalai, G., Kudela, K., Sloan, T., Wolfendale, A.W., 2009,
Some aspects of ionization and the cloud cover, cosmic ray correlation problem, Journal
of Atmospheric and Solar-Terrestrial Physics, doi:10.1016/j.jastp.2009.03.007
\item[] Ermakov, V.I., Bazilevskaya, G.A., Pokrevsky, P.E., Stozhkov, Yu.I.,
Ion balance equation in the atmosphere, 1997, Journal of Geophysical Research, 
$\bf{102}$(D19), 23413.
\item[]  Cotton, W.R. and Anthes, R.A., 1959, 'Storm and Cloud Dynamics', Academic 
Press, London.
\item[] Global Surface Temperature Anomalies:\\
 $http://lwf.ncdc.noaa.gov/oa/climate/research/anomalies/anomalies.html$
\item[] Haigh, J.D., 2007, The Sun and the Earth Climate, Living Reviews in Solar 
Physics., $\bf{4}$, 2.
\item[] Hayakawa S., 1965, 'Cosmic Ray Physics', Monographs and Texts in Physics and 
Astronomy, edited by R.E.Marshak, vol.XXII, John Wiley and Sons, New York, London, 
Sydney, Toronto
\item[] ISCCP data were obtained from the International Satellite Cloud 
Climatology Project web site: {\em http://isccp.giss.nasa.gov}, maintained by the
ISCCP research group at NASA Goddard Institute for Space Science Studied, New York,
Rossow, W.B. and Schiffer, R.A., 1999, 'Advances in understanding clouds from ISCCP', 
Bulletin of the American Meteorological Society, {\bf 80}, 226
\item[] Kirkby, J., 2007, Cosmic Rays and Climate, Surveys in Geophysics, \textbf{28}, 
335
\item[] Kristj\'{a}nsson, J.E., Kristiansen, J., and Kaas, E., 2004, Solar activity, 
cosmic rays and climate - an update, Advances in Space Research, \textbf{34}, 407
\item[] Kristj\'{a}nsson, J.E., Stjern, C.W., Stordal, F., Fj{\ae}raa, A.M., Myhre, G.,
 J\'{o}hansson, K., 2008, Cosmic rays, cloud condensation nuclei and clouds - a 
reassessment using MODIS data, Atmospheric Chemistry and Physics, \textbf{8}, 7373
\item[] Kudryavtsev, I.V. and Yungner, H., 2009, Cosmic rays and variations of the 
concentrations of active nuclei of condensation and crystallization in the Earth's 
atmosphere, Bulletin of the Russian Academy of Sciences, Physics, \textbf{73}, 413
\item[] Marsh, N. and Svensmark, H., 2003, Solar Influence on Earth's Climate,
Space Science Reviews, $\bf{107}$, 317. 
\item[] Mason, B.J., 1971, `The Physics of Clouds' ( Clarendon Press, Oxford ). 
\item[] Palle Bago, E. and Butler, C.J., 2000, The influence of cosmic rays on 
terrestrial clouds and global warming, Astronomy and Geophysics. \textbf{41}, 4.18.
\item[] Sloan, T. and Wolfendale A.W., 2008, Could cosmic rays cause global warming ? 
Environmental Research Letters. $\bf{3}$, 024001.
\item[] Sturrock, P.A., 2008, Solar neutrino variability and its implications for solar
 physics and neutrino physics, arXiv/0810.2755.
\item[] Svensmark, H. and Friis-Christensen, E., 1997, Variation in cosmic ray flux and
 global cloud coverage - a missing link in solar-climate relationship, Journal of 
Atmospheric and Solar-Terrestrial Physics, $\bf{59}$, 1225.
\item[] Svensmark, H., 2007, Cosmoclimatology: a new theory emerges, News and Reviews
in Astronomy and Geophysics, $\bf{48}$, 18
\item[] Uchaikin, V.V. and Ryzhov, V.V., 1988, 'Stochastic theory of transport of high 
energy particles', Nauka, Novosibirsk (~in Russian~).
\item[] Usoskin, I.G., Gladysheva, O.G., Kovaltsov, G.A., 2004, Cosmic ray induced 
ionization in the atmosphere: spatial and temporal changes, Journal of Atmospheric and
 Solar-Terrestrial Physics, $\bf{66}$, 1791.
\item[] WDC neutron data: \\
{\em http://center.stelab.nagoya-u.ac.jp/cawses/cddvd/ob0061.html}
\end{itemize}

\newpage

\begin{center}
\bf{Captions to figures}
\end{center}

Figure 1. The temporal behavior of CR (a), Global LCC (b) and the 
$\chi^2/ndf$ value as a function of the time lag between CR and LCC curves (c). Dotted 
lines in (a) and (b) are the mean values and dashed lines are the 5-degree polynomial 
fits of the CR intensity and LCC respectively. 

Figure 2. Correlation plots for variations of LCC and CR during the 22nd
solar cycle: ~~(a) long-term variations, (b) short-term variations. Dashed lines in 
both panels are linear regression lines, the full line in the (a)-panel is the best fit
 curve of $Y = a + bX^c$ type with coefficients $a,b,c$ indicated inside the panel. The
 slope $b$ of the regression line and the correlation coefficient $r$ are indicated
 inside both panels. 

Figure 3. Left set of panels: the temporal behavior of MCC and 
LCC ( two upper panels ), correlation plot for their long-term and short-term 
variations ( two lower panels ). Right set of panels: the same as the left one, but for
 HCC and MCC respectively. 

Figure 4. Seasonal variation of the Global MCC (a) and LCC (b). 

Figure 5. The seasonal variation of the sea, land and Global temperature 
from (~Global Surface Temperature Anomalies~). Numbers above the curves show the 
fraction of the total area occupied by the  sea (71\%) and land (29\%).

Figure 6. The temporal behavior of CR (a), LCC (b) and the correlation of 
their long-term variations during 22d and 23d cycles of the solar activity (c).

Figure 7. The temporal behavior of the Global surface temperature (a), HCC
 (b), MCC (c) and LCC (d) during 22nd and 23rd cycles of solar activity. Dotted lines
in (b), (c) and (d) panels show mean values of CC for 1985-2005 period of time. The 
dotted line in (a) panel shows the mean temperature during the last century 1900-2000, 
which is equal to $13.86^\circ C$. It illustrates the higher temperature in the last 
two decades of this period - the so called {\em Global Warming}. Dashed lines show the 
linear fits of the temporal behavior of the surface temperature and CC. Full lines are 
their 5-degree polynomial approximations.

Figure 8. The latitude dependence of CC characteristics: (a) absolute 
values of LCC (~ open circles~), MCC (~full circles~) and HCC (~open stars~). (b) LCC, 
MCC and HCC correlations with CR (~Climax~). Notations are the same as in (a). (c) The 
sensitivity of MCC to LCC (~open circles~) and their correlation coefficient (~full 
circles~).

Figure 9. Temporal behavior of the surface temperature in 1880-2005 years 
(~upper panel~) and its Fourier frequency spectrum of the global temperature for this
time range (~lower panel~). The peak at 0.007 month$^{-1}$ corresponds to $\sim$11-year
period coincident with 11.87 year solar cycle (~Sturrock, 2008~).

Figure 10. Temporal behavior of the LCC top pressure, 
fitted by linear (~dotted line~) and 5-degree polynomial (~full line~) approximations.

\newpage 

\begin{figure}[hptb]
\begin{center}
\includegraphics[height=10cm,width=15cm]{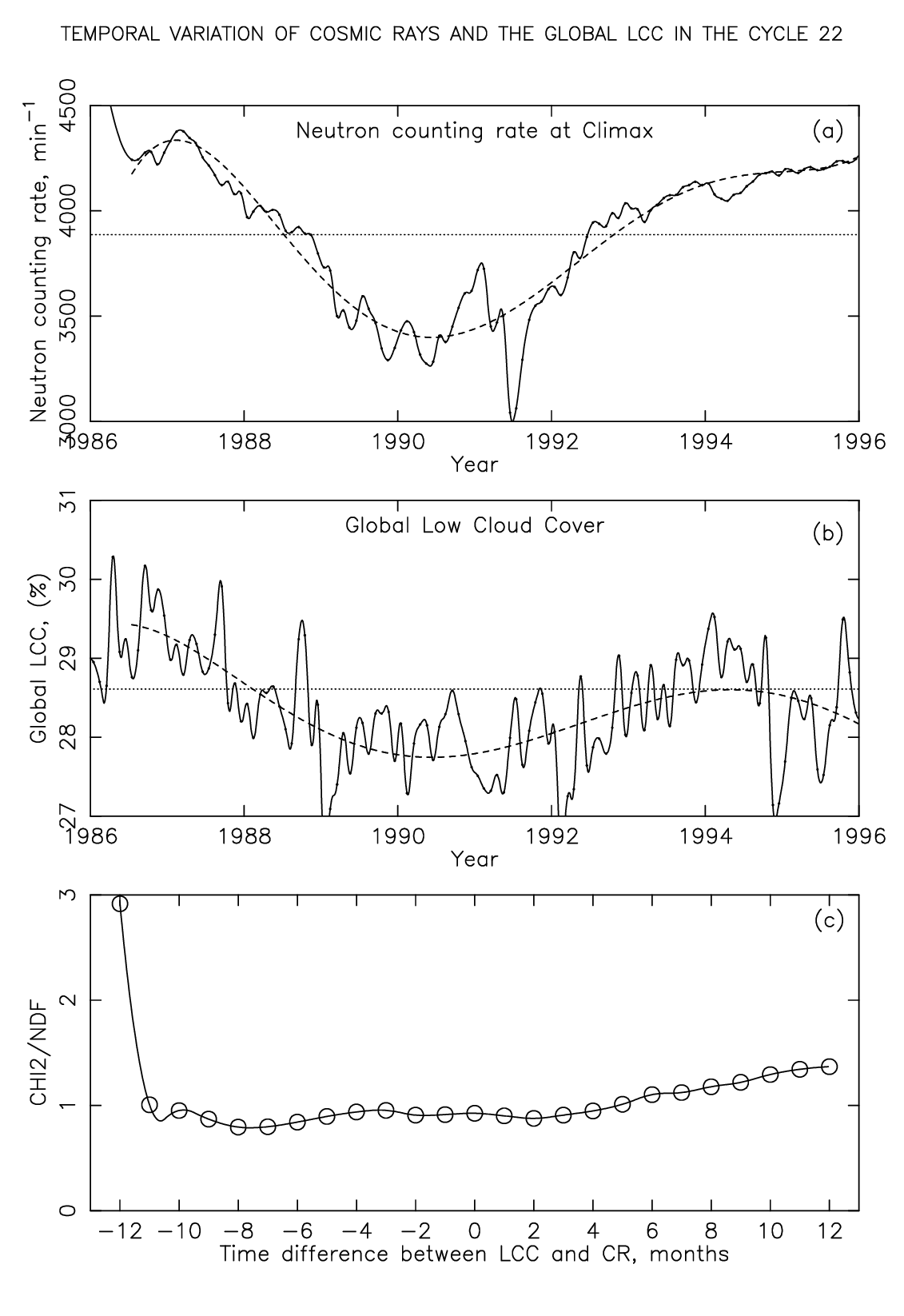}
\caption{}
\label{fig:fig1} 
\end{center}
\end{figure}

\newpage

\begin{figure}[hptb]
\begin{center}
\includegraphics[height=10cm,width=15cm]{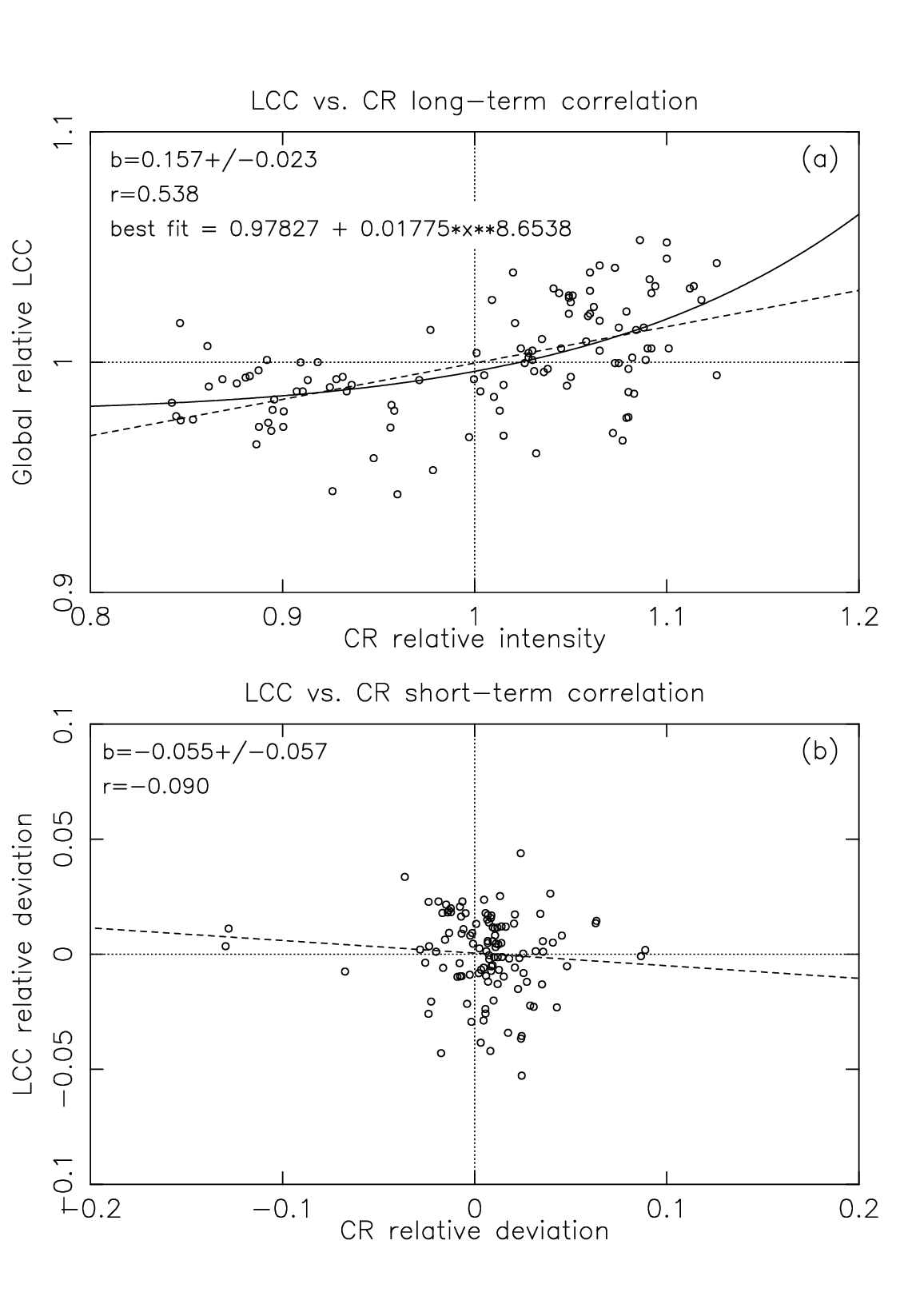}
\caption{} 
\label{fig:fig2} 
\end{center}
\end{figure}

\newpage

\begin{figure}[hptb]
\begin{center}
\includegraphics[height=12cm,width=7.4cm]{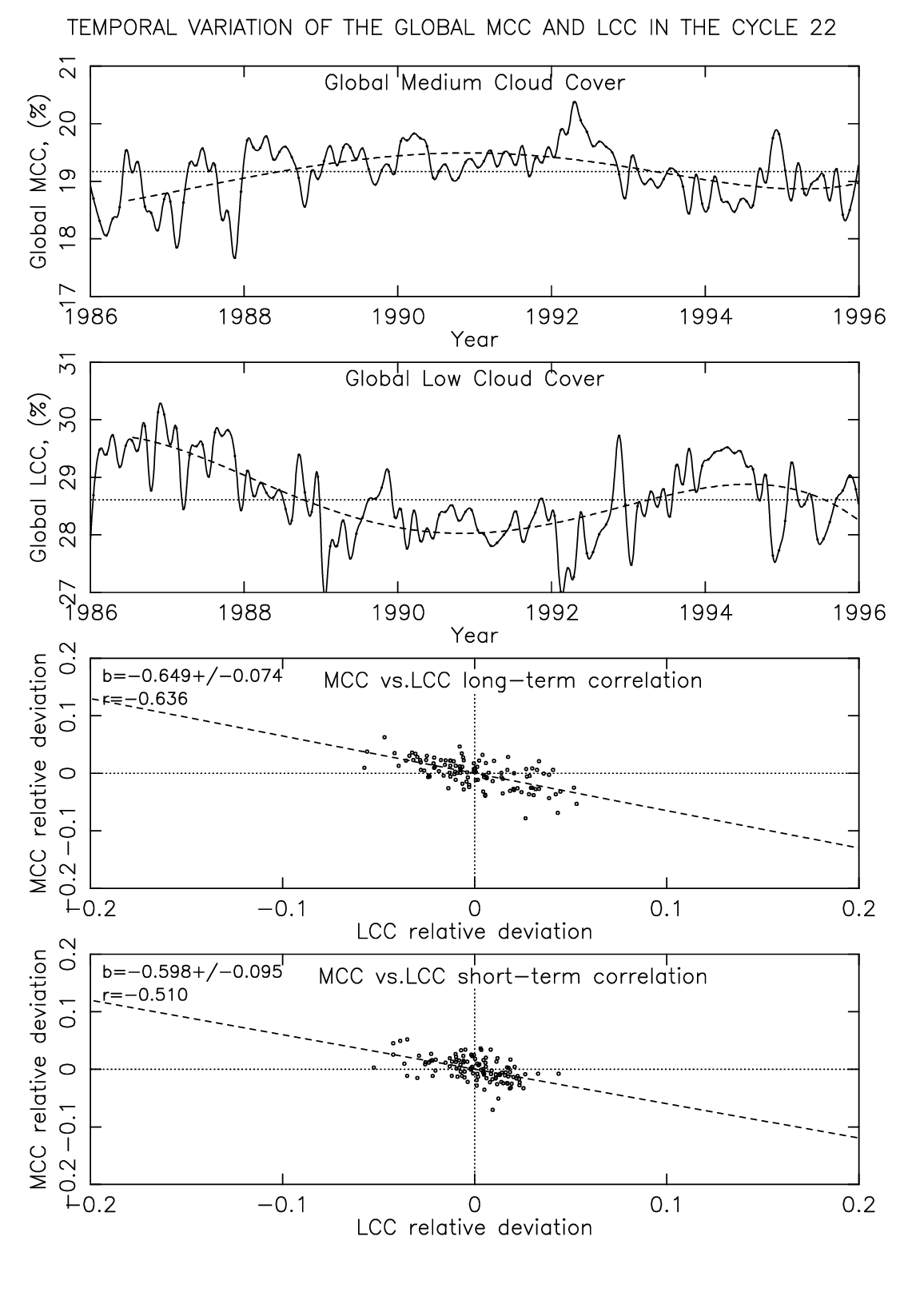}
\includegraphics[height=12cm,width=7.4cm]{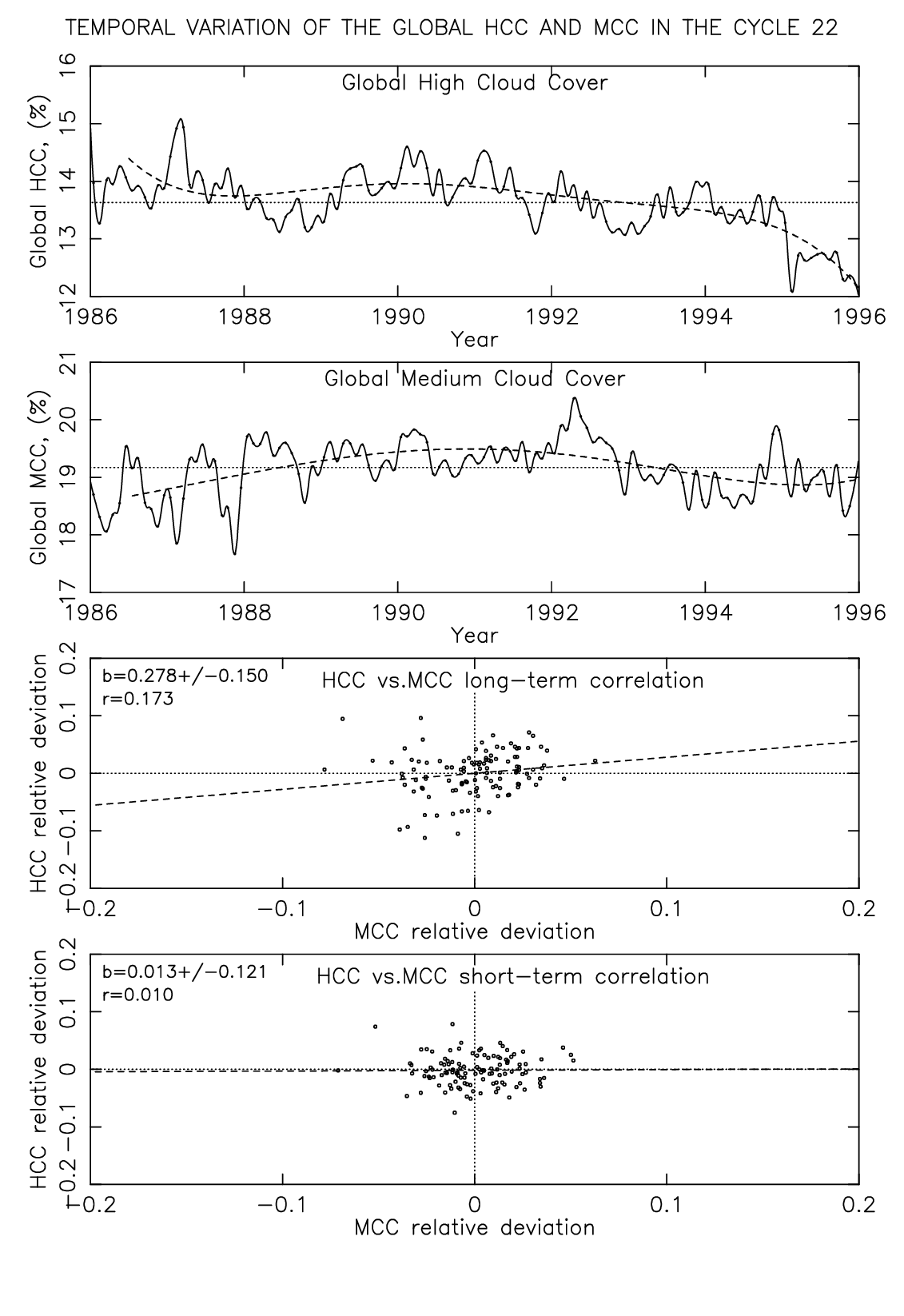}
\caption{} 
\label{fig:fig3} 
\end{center}
\end{figure}                 

\newpage

\begin{figure}[hptb]
\begin{center}
\includegraphics[height=15cm,width=9cm,angle=-90]{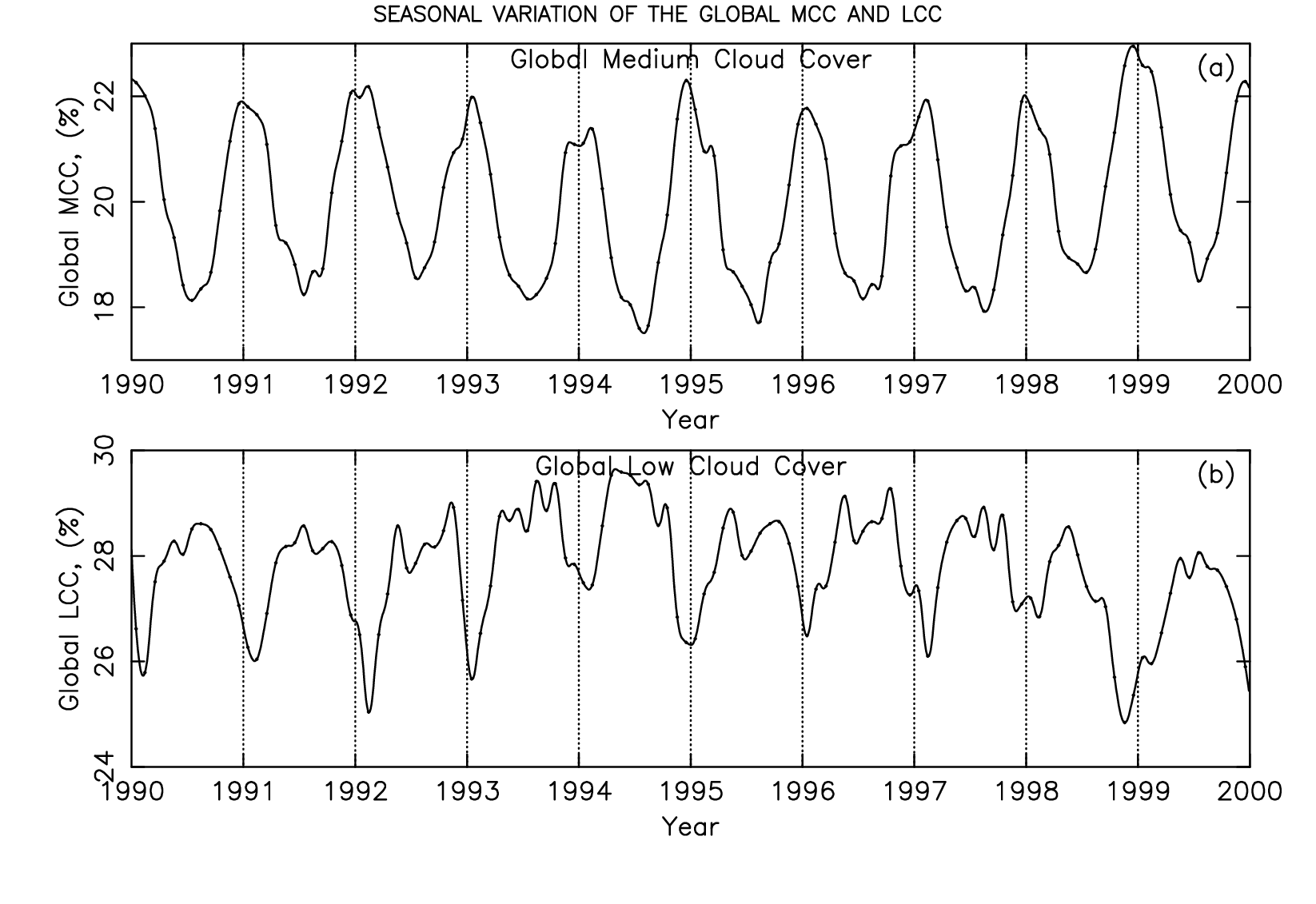}
\caption{} 
\label{fig:fig4} 
\end{center}
\end{figure}

\newpage

\begin{figure}[hptb]
\begin{center}
\includegraphics[height=15cm,width=8cm,angle=-90]{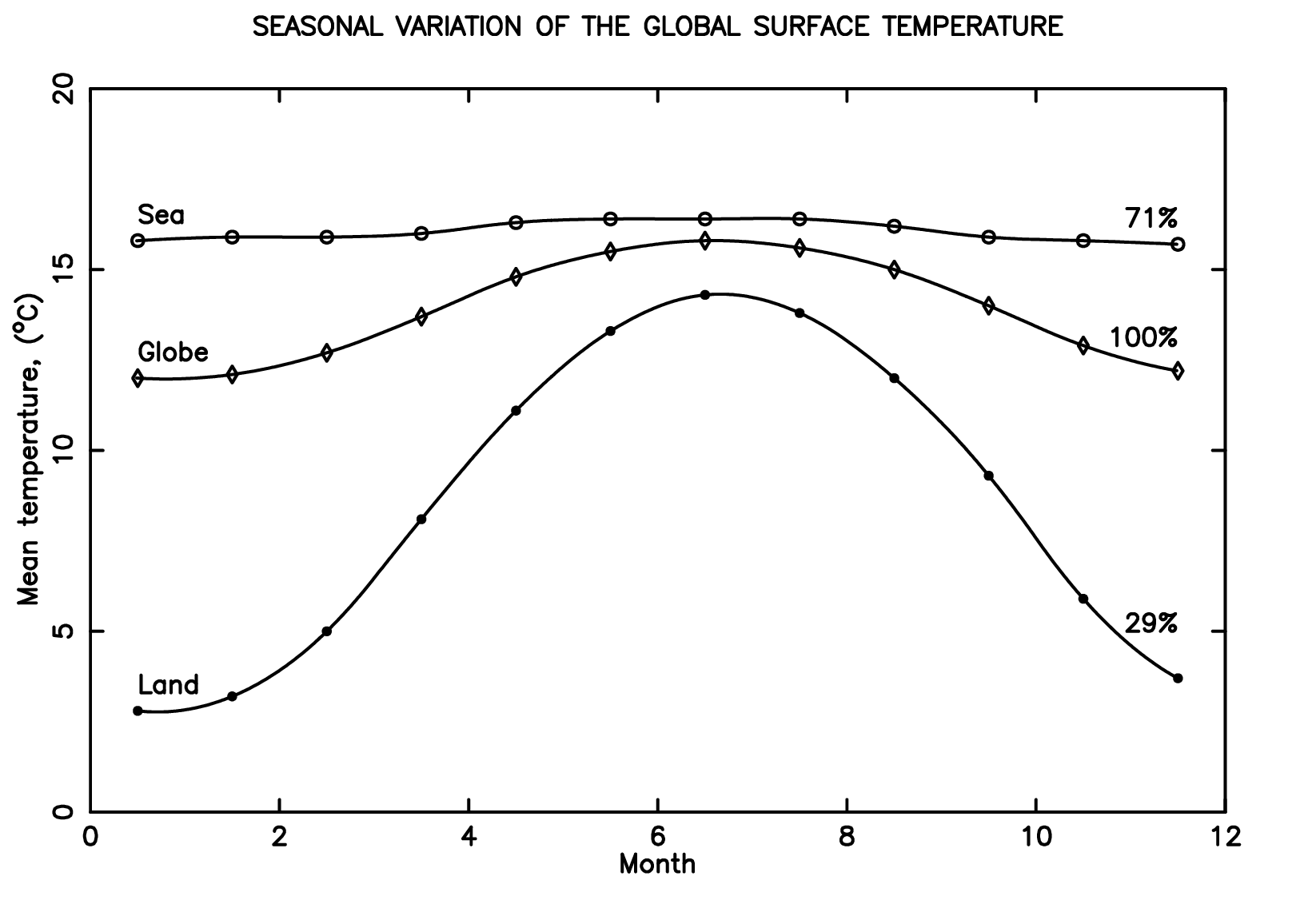}
\caption{\footnotesize}
\label{fig:fig5} 
\end{center}
\end{figure} 

\newpage

\begin{figure}[hptb]
\begin{center}
\includegraphics[height=12cm,width=15cm]{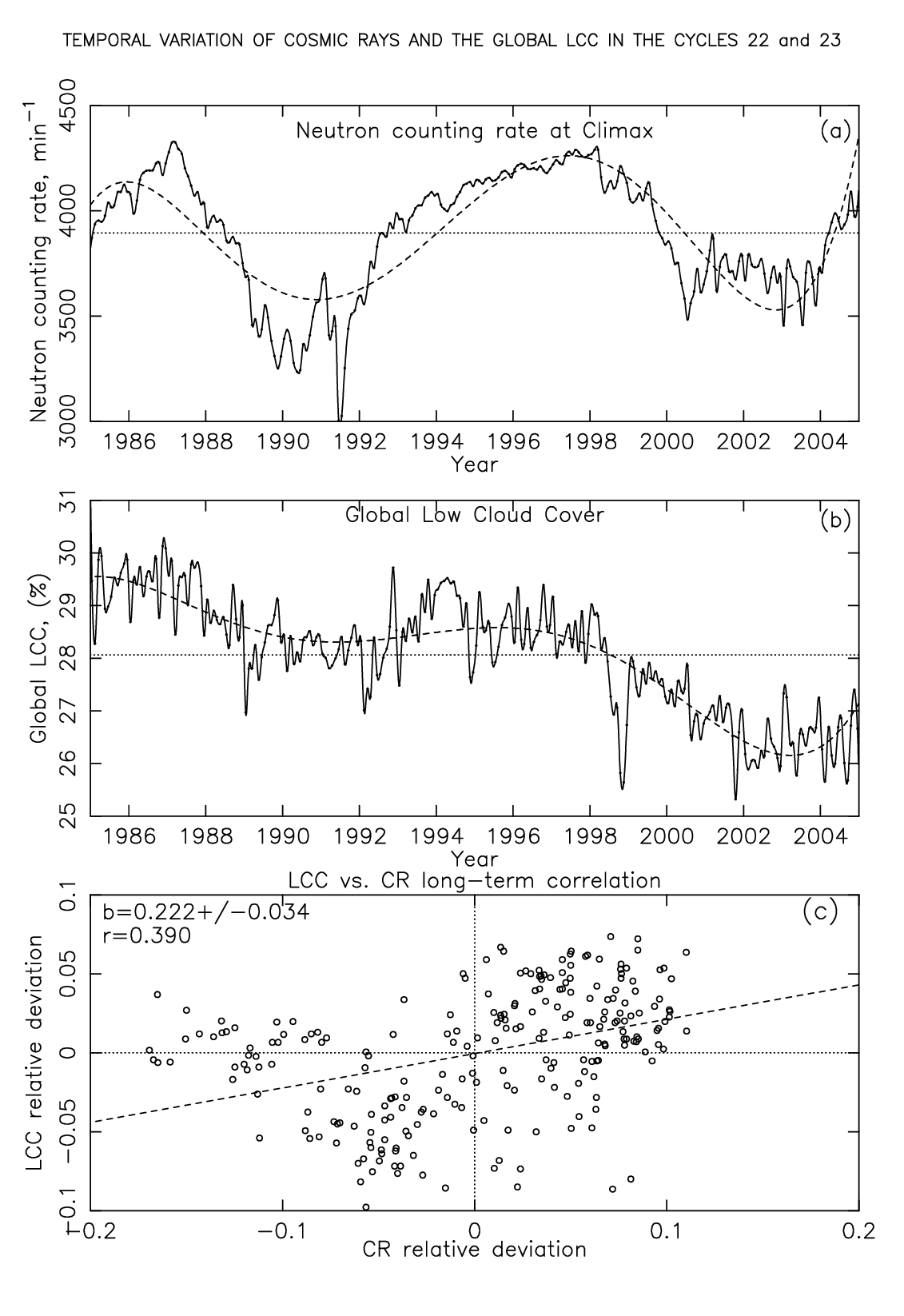}
\caption{\footnotesize} 
\label{fig:fig6} 
\end{center}
\end{figure}

\newpage

\begin{figure}[hptb]
\begin{center}
\includegraphics[height=15cm,width=15cm]{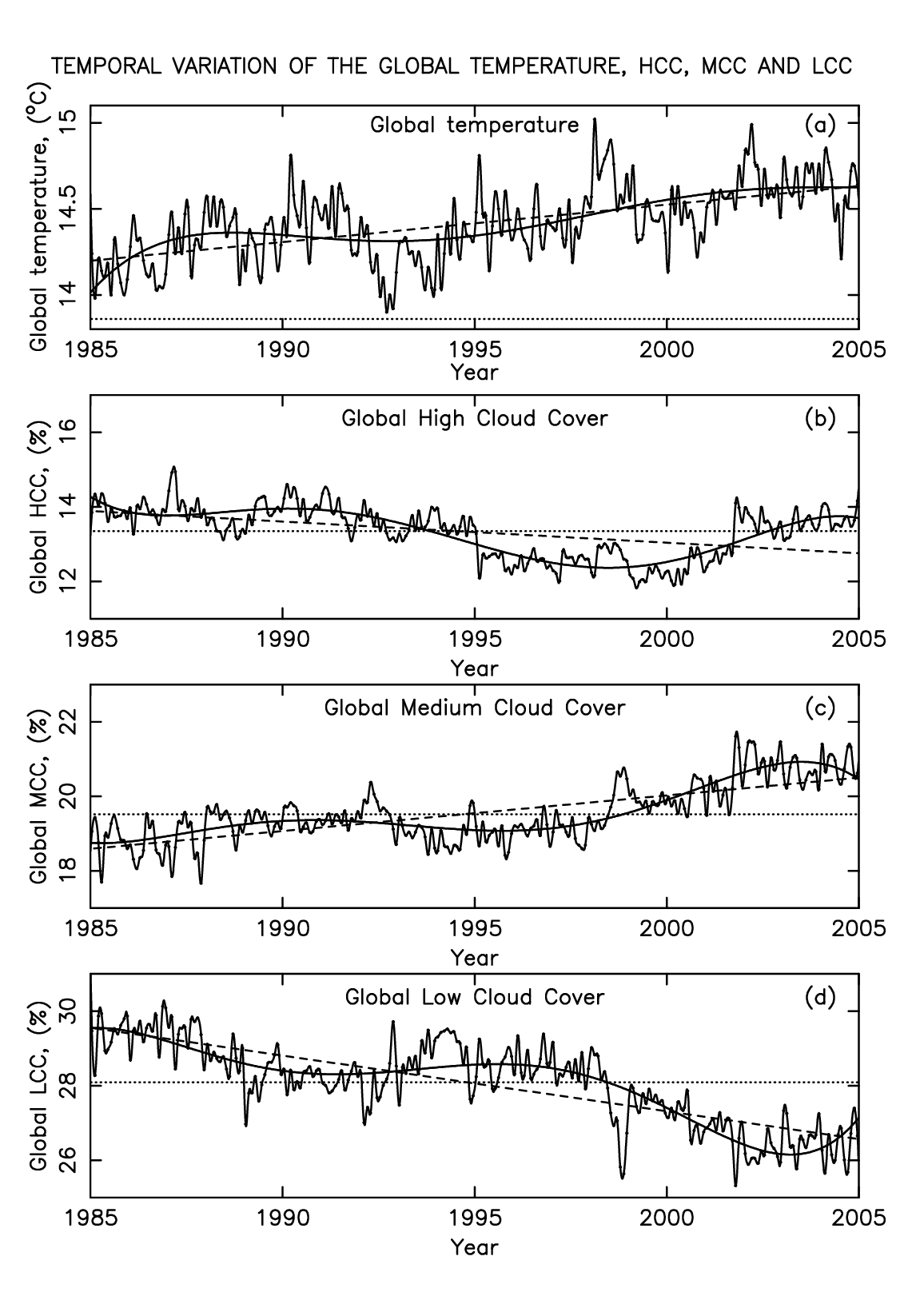}
\caption{\footnotesize}.
\label{fig:fig7} 
\end{center}
\end{figure}

\newpage

\begin{figure}[hptt]
\begin{center}
\includegraphics[height=12cm,width=15cm]{fig8.eps}
\caption{\footnotesize} 
\label{fig:fig8} 
\end{center}
\end{figure}

\newpage

\begin{figure}[hptb]
\begin{center}
\includegraphics[height=15cm,width=15cm]{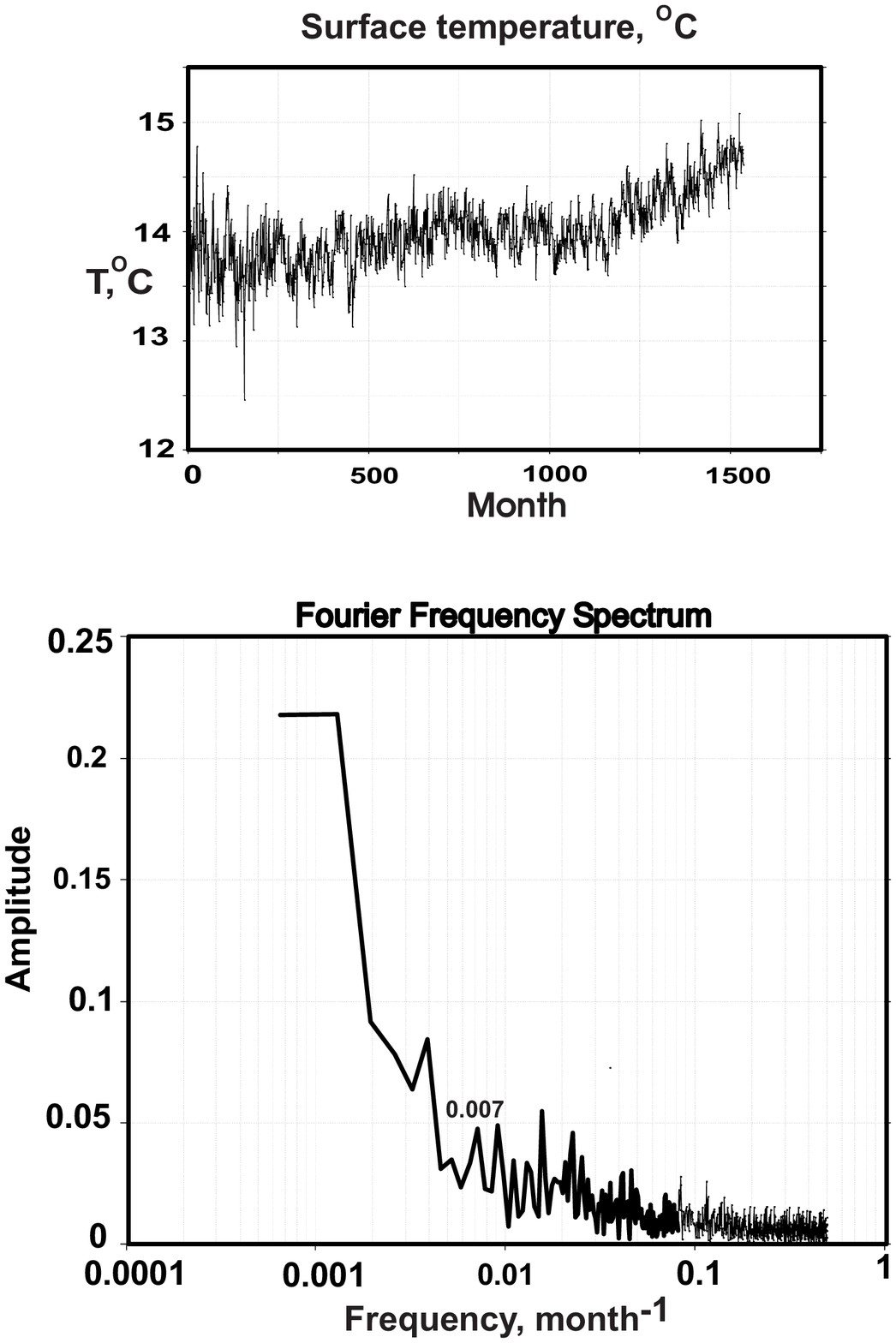}
\caption{\footnotesize}
\label{fig:fig9} 
\end{center}
\end{figure}

\newpage

\begin{figure}[hptb]
\begin{center}
\includegraphics[height=15cm,width=12cm,angle=-90]{fig10.eps}
\caption{\footnotesize}  
\label{fig:fig10} 
\end{center}
\end{figure}

\end{document}